\documentclass[submission,copyright,creativecommons]{eptcs}
\providecommand{\event}{LSFA 2022}

\usepackage{amssymb,amsmath,amsthm}
\usepackage{bbm}
\usepackage{subfigure}
\usepackage{tikz}
\usepackage{enumerate}

\usepackage{epsf}
\usepackage[dvips]{epsfig,graphicx}
\usepackage{subfigure}
\usepackage{float}
\usepackage[utf8x]{inputenc}

\usepackage{latexsym}
\usepackage{textcomp}

\usepackage{amsmath}
\usepackage{amssymb}
\usepackage{amsthm}
\usepackage{mathrsfs}

\usepackage{url}
\usepackage{longtable}
\usepackage{lscape}

\usepackage{colortbl}
\usepackage[boxruled,longend,linesnumbered,norelsize]{algorithm2e}

\usepackage{xcolor}

\usepackage{multicol}
\usepackage{multirow}
\usepackage{rotating}
\usepackage{listings}

\usepackage{enumitem}

\usepackage{amsthm}
\usepackage{tikz}
\usetikzlibrary{shapes.geometric}
\usetikzlibrary{topaths,calc}
\usepackage{tikz-qtree}
\usetikzlibrary{fit}
\usetikzlibrary{automata, positioning, arrows}
\usepackage{multicol}
\theoremstyle{definition}
\newtheorem{definition}{Definition}
\theoremstyle{definition}
\newtheorem{example}{Example}
\theoremstyle{theorem}
\newtheorem{theorem}{Theorem}
\theoremstyle{lemma}
\newtheorem{lemma}{Lemma}
\usepackage{xcolor}
\usepackage{syntax}
\usepackage{newfloat}
\usepackage{blindtext}
\usepackage[final]{pdfpages}
\usepackage{listings}
\usepackage{fancyvrb}
\usepackage{reo}
\usepackage{varwidth}
\usepackage{enumitem}
\usepackage{bussproofs}
\usepackage{prooftrees}
\usepackage[color]{coqdoc}
\usepackage{booktabs}

\usepackage{epsf}
\usepackage[dvips]{epsfig,graphicx}
\usepackage{subfigure}
\usepackage{float}

\usepackage{latexsym}
\usepackage{textcomp}

\usepackage{amsmath}
\usepackage{amssymb}
\usepackage{amsthm}
\usepackage{mathrsfs}

\usepackage{url}
\usepackage{longtable}
\usepackage{lscape}

\usepackage{multicol}
\usepackage{multirow}
\usepackage{rotating}
\usepackage{listings}

\usepackage{xcolor}
\usepackage{syntax}
\usepackage{newfloat}
\usepackage{blindtext}
\usepackage[final]{pdfpages}
\usepackage{listings}
\usepackage{fancyvrb}
\usepackage{reo}
\usepackage{varwidth}
\usepackage{enumitem}
\usepackage{bussproofs}
\usepackage{prooftrees}
\usepackage[color]{coqdoc}
\usepackage{booktabs}
\usepackage[boxruled,longend,linesnumbered,norelsize]{algorithm2e}

\usepackage{enumitem}

\usetikzlibrary{arrows,shapes,decorations,petri}
\tikzset{every node/.style={scale=.71}}

\newcommand{\relo}{$ReLo$}

\title{\relo: a Dynamic Logic to Reason About Reo Circuits\thanks{This work was supported by CNPq and FAPERJ.}}
\author{Erick Grilo
\institute{Instituto de Computa\c c\~ao}
\institute{Universidade Federal Fluminense}
\email{simas\_grilo@id.uff.br}
\and
Bruno Lopes
\institute{Instituto de Computa\c c\~ao}
\institute{Universidade Federal Fluminense}
\email{bruno@ic.uff.br}
}

\def\titlerunning{\relo\: a Dynamic Logic to Reason About Reo Circuits}
\def\authorrunning{E.\ Grilo \& B.\ Lopes}

\begin{document}

\maketitle

\begin{abstract}
	
Critical systems require high reliability and are present in many domains. They are systems in which failure may result in financial damage or even loss of lives. Standard
techniques of software engineering are not enough to ensure the absence of unacceptable
failures and/or that critical requirements are fulfilled.
Reo is a component-based modelling language that aims to provide a framework to build software based on existing pieces of software, which has been used in a wide variety of domains. Its formal semantics provides grounds to certify that systems based on Reo models satisfy specific requirements (i.e., absence of deadlocks). 
Current logical approaches for reasoning over Reo require the conversion of formal semantics into a logical framework. \relo\ is a dynamic logic that naturally subsumes Reo's semantics. It provides a means to reason over Reo circuits.
This work extends \relo\ by introducing the iteration operator, and soundness and completeness proofs for its axiomatization.
The core aspects of this logic are also formalized in the Coq proof assistant. 
\end{abstract}

\section{Introduction}

In software development, service-oriented computing~\cite{Papazoglou2003} and model-driven development~\cite{Atkinson2003} are examples of techniques that take advantage of software models. The first technique advocates computing based on preexisting systems (services) as described by Service-Oriented Architecture (SOA), while the latter is a development technique that considers the implementation of a system based on a model. A model is an abstraction of a system (or some particular portion of it) in a specific language, which will be used as a specification basis for the system's implementation. It can be specified in languages such as Unified Modeling Language (UML) or formal specification languages like B~\cite{Abrial1991} or Alloy~\cite{Jackson2002}. Researchers also have applied approaches such as formal methods in software development to formalize and assure that certain (critical) systems have some required properties~\cite{Knight2002,Ostro1992}.

Reo~\cite{Arbab2004} is a prominent modelling language, enabling coordination of communication between interconnected systems without focusing on their internal properties. Reo models are compositionally built from base connectors, where each connector in Reo stands for a specific communication pattern. Reo has proven to be successful in modeling the organization of concurrent systems' interaction, being used in a variety of applications, from process modeling to Web-Services integration~\cite{Arbab2008A} and even in the construction of frameworks to verify specifications in Reo~\cite{Kokash2012,Tasharofi2009}.

Reo's ability to model communication between software interfaces has also attracted research on verification of Reo circuits, resulting in many different formal semantics~\cite{Jongmans2012} like automata-based models~\cite{Arbab2006,Baier2005,Kokash2010}, coalgebraic models~\cite{Arbab2004}, Intuitionistic Logic with Petri Nets~\cite{Clarke2007} (to name a few), and some of their implementations~\cite{Kokash2012,Sun2014,Li2016,Li2017,Li2018,Zhang2019,Kokash2010}. However, as far as the authors are concerned, there is no logic apart from \relo~\cite{Grilo2020-2} to specific reason about Reo models naturally, where the usage of other logic-based approaches requires conversion between different formal semantics.

This work extends \relo~\cite{Grilo2020-2} by introducing an iteration operator and the soundness and completeness proofs of its axiomatic system. A prototypical implementation of this framework in Coq proof assistant, enabling the verification of properties of Reo programs in \relo\ within a computerized environment is available at \url{http://github.com/frame-lab/ReoLogicCoq}.

This work is structured as follows. Section~\ref{sec:rlf} discusses briefly a related logic formalism with the one hereby proposed and introduces Reo modelling language, along with some examples. 
 Section~\ref{sec:reloDef} discuss \relo's main aspects, from its core definitions (such as language, models, transitions firing) and its soundness and completeness proofs.
 Finally, Section~\ref{sec:conclusions} closes the work by discussing the obtained results and assessing possible future work.

\section{Related Work} \label{sec:related}

The fact that Reo can be used to model many real-world situations has attracted attention from researchers all around the world, resulting in a great effort directed in formalizing means to verify properties of Reo models~\cite{Klein2011,Pourvatan2009,Kokash2011,Kokash2012,Mousavi2006,Li2019,Jongmans2012}. Such effort also resulted in the proposal of several formal semantics for this modelling language~\cite{Jongmans2012}, varying from operational semantics to coalgebric models.

One of the most known formal semantics for Reo consists of Constraint Automata~\cite{Baier2006}, an operational semantic in which Reo connectors are modelled as automata for $TDS$-languages~\cite{Arbab2002A}. It enables reasoning over the data flow of Reo connectors and when they happened. Constraint Automata have been extended to some variants which aim to enrich the reasoning process by capturing properties like the timing of the data flows or possible actions over the data, respectively as Timed Constraint Automata~\cite{Kokash2012} and Action Constraint Automata~\cite{Kokash2010-2}. Some of them are briefly discussed below, along with other formal semantics for Reo.

The approach presented by Klein et al.~\cite{Klein2011} provides a platform to reason about Reo models using Vereofy,\footnote{\url{http://www.vereofy.de}} a model checker for component-based systems, while Pourvatan et al.~\cite{Pourvatan2009} propose an approach to reason about Reo models employing symbolic execution of Constraint Automata. Kokash \& Arbab~\cite{Kokash2011} formally verify Long-Running Transactions (LRTs) modelled as Reo connectors using Vereofy, enabling expressing properties of these connectors in logics such as Linear Temporal Logic (LTL) or a variant of Computation Tree Logic (CTL) named Alternating-time Stream Logic (ASL). Kokash et al.~\cite{Kokash2012} use mCRL2 to encode Reo's semantics in Constraint Automata and other automata-based semantics, encoding their behaviour as mCRL2 processes and enabling the expression of properties regarding deadlocks and data constraints which depend upon time. mCRL2 also supports model-checking of Reo in a dynamic logic (with fixed points), where modalities are regular expressions, atomic actions are sets of nodes that fire at the same time. Mouzavi et al.~\cite{Mousavi2006} propose an approach based on Maude to model checking Reo models, encoding Reo's operational semantics of the connectors. 

Proof assistants have been used to reason about Reo connectors~\cite{Li2015,Li2017,Li2018,Li2016,Zhang2019,REOJAL}. The approaches adopted by Li et al.~\cite{Li2015,Li2016,REOJAL} are among the ones that employ Coq to verify Reo models formally. In~\cite{Li2015} a formalization of four of the Reo canonical connectors (Sync, FIFO1, SyncDrain, and LossySync) along with an LTL-based language defined as an inductive type in Coq is presented, while~\cite{Li2016} proposes the formalization of five Reo canonical channels
and a procedure that creates composite channels by logical conjunction of the connectors modelled. 

In~\cite{REOJAL}, a framework to provide means of graphically model Reo connectors and validate the generated model in Constraint Automata using Coq and NuSMV\footnote{\url{https://nusmv.fbk.eu/}} is discussed. It also enables the automatic generation of Coq code to a Haskell model employing the Coq code extraction apparatus.
%
When restricting the works considering logics and Reo, as far as the authors know there is only the work by~\cite{Clarke2007} which focuses on formalizing the semantics of Reo connectors Sync, LossySync, FIFO1, SyncDrain, AsyncDrain, Filter, Transform, Merger, and Replicator in terms of zero-safe Petri nets~\cite{Bruni2000}, a special class of Petri-nets with two types of places: zero and stable places. This encoding is then converted to terms in Intuitionistic Temporal Linear Logic, enabling reasoning about Reo connectors in this logic.


%

\section{Background} \label{sec:rlf}

This section provides a succinct overview of Reo~\cite{Arbab2004,Arbab2006}, considering its main characteristics and a modelling examples as it is the target language \relo\ provides a formal semantic to reason over. 

\subsection{The Reo Modelling Language} \label{sec:reo}

As a coordination model, Reo focuses on connectors, their composition, and how they behave, not focusing on particular details regarding the entities that are connected, communicate, and interact through those connectors. Connected entities may be modules of sequential code, objects, agents, processes, web services, and any other software component where its integration with other software can be used to build a system~\cite{Arbab2004}. Such entities are defined as component instances in Reo.

Channels in Reo are defined as a point-to-point link between two distinct nodes, where each channel has its unique predefined behavior. Each channel in Reo has exactly two ends, which can be of the following types: the source end, which accepts data into the channel, and the sink end, which dispenses data out of the channel. Channels are used to compose more complex connectors, being possible to combine user-defined channels amongst themselves and with the canonical connectors provided by Baier et al.~\cite{Baier2006}. Figure~\ref{fig:reoCanonical} shows the basic set of connectors as presented by Kokash et al.~\cite{Kokash2012}. 

\usetikzlibrary{snakes}	
\begin{figure*}[!htb] 
	\centering\subfigure[Sync]{	
		\centering \begin{tikzpicture}[->,>=stealth',font=\sffamily,semithick,node distance=1.8cm]
		\node (A)  {$A$};
		\node (B) [right of=A]  {$B$};
		\path (A) edge  node {} (B);
		
		\end{tikzpicture}
	}\quad
	\subfigure[LossySync]{
		\centering \begin{tikzpicture}[->,>=stealth',font=\sffamily,semithick,node distance=1.8cm,dashed]
		\node (A)  {$A$};
		\node (B) [right of=A]  {$B$};
		\path (A) edge  node {} (B);
		
		\end{tikzpicture}
	}\quad
	\subfigure[FIFO]{
		\centering \begin{tikzpicture}[->,>=stealth,font=\sffamily,semithick,node distance=1cm]
		\node (A)  {$A$};
		\node (x) [draw,rectangle,minimum width=0.5cm,minimum height=0.2cm] at  (.95,0) {}; 
		\node (B) [right of=x]  {$B$};
		\draw[semithick,-] (A) -- ++(.7cm,0) |- (.5,0);
		\draw [semithick,->](x) edge  node {} (B);
		
		\end{tikzpicture}	
	}
	\subfigure[SyncDrain]{
		\centering \begin{tikzpicture}[->,>=to reversed,font=\sffamily,semithick,node distance=1.8cm]
		\node (A)  {$A$};
		\node (B) [right of=A]  {$B$};
		\path (A) edge  node {} (B)
		(B) edge  node {} (A);
		
		\end{tikzpicture}	
	}\quad
	\subfigure[AsyncDrain]{
		\centering \begin{tikzpicture}[->,>=to reversed,font=\sffamily,semithick,node distance=1.8cm]
		\node (A)  {$A$};
		\node (B) [right of=A]  {$B$};
		\path (A) edge  node {\tikz \draw[|-|,semithick ] (0,0) -- +(.1,0);} (B)
		(B) edge  node {} (A);
		
		\end{tikzpicture}	
	}
	
	\subfigure[\mbox{Filter}]{
		\centering \begin{tikzpicture}[->,decorate,decoration=snake,>=stealth,font=\sffamily,semithick,node distance=1.8cm]
		\node (A)  {$A$};
		\node (B) [right of=A]  {$B$};
		\draw [->,decorate,decoration=snake] (A) -- (B);
		\end{tikzpicture}	
	}\quad
	\subfigure[\mbox{Transform}]{
		\centering \begin{tikzpicture}[->,>=stealth,font=\sffamily,semithick,node distance=1.8cm,]
		\node (A)  {$A$};
		\node (B) [right of=A]  {$B$};
		\path (A) edge node {\tikz \draw[-triangle 90,] (0,0) -- +(.1,0);} (B);
		
		\end{tikzpicture}	
	}\quad
	\subfigure[Merger]{
		\centering \begin{tikzpicture}[>=stealth,font=\sffamily,semithick,node distance=.8cm]
		\node (A) at (0,0) {$A$};
		\node (B) at (0,0.8)  {$B$};
		\node (x) [fill,circle,inner sep=1pt] at (0.8,0.4) {};
		\node (C) [right of=x] {$C$};
		\path (A) edge  node {} (x)
		(B) edge  node {} (x)
		(x) edge  node {} (C); 
		
		\end{tikzpicture}	
	}\quad
	\subfigure[\mbox{Replicator}]{
		\centering \begin{tikzpicture}[>=stealth,font=\sffamily,semithick,node distance=1cm]
		\node (A) {$A$};
		\node (x) [fill,circle,inner sep=1pt] at (0.8,0) {}; 
		\node (B) at (1.6,-0.4) {$B$};
		\node (C) at (1.6,0.4)  {$C$};
		\path (A) edge  node {} (x)
		(x) edge  node {} (B)
		(x) edge node  {} (C);
		
		\end{tikzpicture}	
	}
	\caption{Canonical Reo connectors}
	\label{fig:reoCanonical}
\end{figure*}
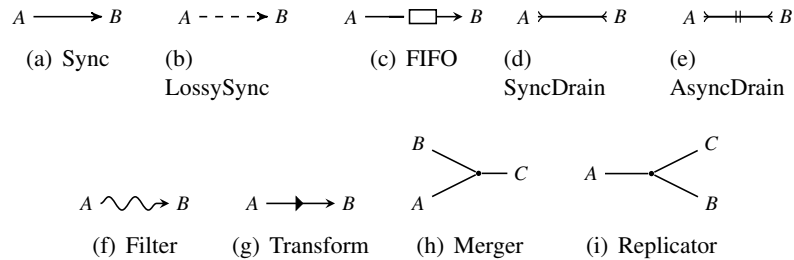

Channel ends can be used by any entity to send/receive data, given that the entity belongs to an instance that knows these ends. Entities may use channels only if the instance they belong to is connected to one of the channel ends, enabling either sending or receiving data (depending on the kind of channel end the entity has access to).

The bound between a software instance and a channel end is a logical connection that does not rely on properties such as the location of the involved entities. Channels in Reo have the sole objective to enable the data exchange following the behaviour of the connectors composing the channel, utilizing I/O operations predefined for each entity in an instance. A channel can be known by zero or more instances at a time, but its ends can be used by at most one entity at the same time. 



\begin{figure}[!htb] 
	\centering 
		\begin{tikzpicture}[>=stealth,font=\sffamily,semithick,node distance=1.6cm]
		\node (bo1) [draw,scale=0.6,fill,circle,color=green!70!black] at (1.2,-1.5) {};
		\node (A) [] at (0.4,-1.5) {$X$};
		\node (y) [] at (2,-1.5) {$Y$};
		\node (B) [above of = y] {$A$};
		\node (x) [draw,rectangle,minimum width=0.5cm,minimum height=0.2cm] at (1.2,-1.5) {};
		\node (z) [draw,rectangle,minimum width=0.5cm,minimum height=0.2cm] at (2.8,-1.5) {};
		\node (a) [] at (3.6,-1.5) {$W$};
		\node (C) [above of = a] {$B$};
		\node (b) [] at (5.2,-1.5) {$Z$};
		\node (c) [draw,rectangle,minimum width=0.5cm,minimum height=0.2cm] at (4.4,-1.5) {};
		\node (D) [above of = b] {$C$};
		\draw[semithick,-] (b) -- ++(0,-0.5) |- (5.2,-2.5);
		\draw[semithick,-] (5.2,-2.5) -- ++(0,0) |- (0.4,-2.5);
		\draw[semithick,->] (0.4,-2.5) |- (0.4,-1.9);	
		\path (A) edge[-] node{} (x)
		(x) edge[->] node{} (y)
		(y) edge[->] node{} (B)
		(y) edge[-] node{} (z)
		(z) edge[->] node{} (a)
		(a) edge[->] node{} (C)
		(a) edge[-] node{} (c)
		(c) edge[->] node{} (b)
		(b) edge[->] node{} (D);
		\end{tikzpicture}
	\caption{Modelling of the Sequencer in Reo}
	\label{fig:SequencerReo}
\end{figure}
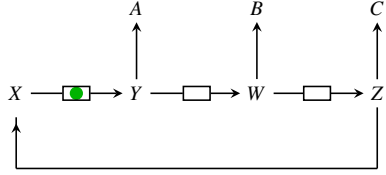

Figure~\ref{fig:SequencerReo} introduces a Reo connector known as Sequencer\footnote{\url{http://arcatools.org/reo}}. It models the data flow between three entities sequentially. The data flows from the first FIFO connector (a buffer), which will be sequentially synchronized with entities in port names names A, B, and C. The Sequencer can be used to model scenarios where processes sequentially interact between themselves.

In short, Reo circuits may be understood as data flowing from different interfaces (i.e., port names connected to a node), where the connector itself models the communication pattern between two of these interfaces. A \relo\ program is composed of one or more Reo connectors as introduced in Figure~\ref{fig:reoCanonical}.

\section{A \relo\ Primer}\label{sec:reloDef}

\relo~\cite{Grilo2020-2} was tailored to subsume Reo models' behaviour naturally in a logic, without needing any mechanism to convert a Reo model denoted by one of its formal semantics to some logical framework. Each basic Reo connector is modelled in the logic's language, which is defined as follows.

\begin{definition}[\relo's language] \label{def:reloLanguage} The language of \relo\ consists of the following:
	
	\begin{itemize}[noitemsep]
		\item An enumerable set of propositions $\Phi$.

		\item Reo channels as denoted by Figure~\ref{fig:reoCanonical}
		
		\item A set of port names $ \mathcal{N}$ 
		
		\item A sequence $Seq_\Pi = \{\epsilon, s_1, s_2, \dots\}$ of data flows in ports of a \relo\ program $\Pi$ (defined below). We define $s_i \leq s_j$ if $s_i$ is a proper (i.e., $s_j$ contains all of $s_i$'s data). Each sequence $s_i$ denotes the data flow of the Reo program $\Pi$ (i.e., all ports that have data synchronized at a specific moment in time) and $\epsilon$ is the empty sequence
		
		\item Program composition symbol : $\odot$
		\item A sequence $t$ of data flows of ports $p$ with data values \{0,1\}, which denotes whether $p$ contains a data item. This describes a data flow occurring in the Reo channel. A BNF describing $t$ is defined as follows:
		\begin{grammar}
			<t> ::= <portName> <data> , <t> | <data> <portName> <data> , <t> \\ | <data> <portName> <data> | <portName> <data> \\
			<portName> ::= $p \in \mathcal{N}$\\
			<data> ::= 0 | 1
		\end{grammar}
		
		\item Iteration operator $^\star$
		
		
	\end{itemize}
\end{definition}

A \relo\ program is defined as any Reo model built from the composition of Reo channels $\pi_i$. In \relo\, their composition is $\Pi = (f,b)$, $\Pi = \pi_1 \odot \pi_2 \odot \dots \odot \pi_n$, and $\pi_i = (f_1, b_1)$. $\odot$ follows the same notion of Reo composition, by ``gluing'' sink nodes of a connector to the source nodes of the other connector.

The set $f$ is the set of connectors $p$ of the model where data flows in and out of the channel (the connector has at least a source node and a sink node), namely Sync, LossySync, FIFO, Filter, Transform, Merger and Replicator. The set $b$ is the set of blocking channels (channels without sink nodes whose inability to fire prevents the remainder of connectors related to their port names from fire), namely SyncDrain and AsyncDrain. 

The following is a simple yet intuitive example of the structure of data flows in \relo. Let the sequence $t$ be $t = \{ A1, B1C\}$. It states that the port $A$ has the data item $1$ in its current data flow, while there is a data item $1$ in the FIFO between $B$ and $C$.

\begin{definition}[\relo\ formulae] \label{def:logicalFormulaRelo} \mbox{}\\
	We define formulae in \relo\ as follows: $\phi = p \mid \top \mid \neg\phi \mid \phi \land \psi \mid \langle t, \pi \rangle \phi$, such that $p\in\Phi$.
	We use the standard abbreviations $\top \equiv \neg \bot, \phi \lor \psi \equiv \neg(\neg \phi \land \neg \psi), \phi \to \psi \equiv \neg\phi \lor \psi$ and $\lbrack t, \pi \rbrack \varphi \equiv \neg \langle t,\pi \rangle \neg \phi$, where $\pi$ is some Reo program and $t$ a data flow.
	
\end{definition}


The connectors in Figure~\ref{fig:reoModels} exemplify compound Reo connectors. 
The model SyncFIFO is composed of a FIFO and a Sync connector in which the data leaving the FIFO is sent to $C$ from $B$ synchronously. Suppose that there is data in the FIFO and in port $B$ ($t = \{A1B,B0\}$). If the FIFO from $A$ to $B$ is processed first then the Sync between $B$ and $C$, the data flow in $B$ will be overwritten before it is sent to $C$, which is not the correct behaviour. The Sync from $B$ to $C$ must fire before the FIFO from $A$ to $B$.

Another example is denoted by the model Sync2Drain. Suppose there is data only in port name $A$ ($t = \{A1\}$). If the Sync from $B$ to $A$ is evaluated first then the SyncDrain between $B$ and $C$, the restriction imposed by the fact that the condition required for the SyncDrain to fire was not met (as $C$'s data flow differs from $B$'s at this moment) is not considered, and data will wrongly flow from $B$ to $A$. The SyncDrain must be first evaluated before all flows as they may block the flow from data of its ports to other channels.

\usetikzlibrary{snakes}	
\begin{figure*}[!htb] 
	\centering\subfigure[SyncFIFO]{	
		\centering \begin{tikzpicture}[->,>=stealth',font=\sffamily,semithick,node distance=1.8cm]
		\node (A)  {$A$};
		\node (x) [draw,rectangle,minimum width=0.5cm,minimum height=0.2cm] at  (.6,0) {};
		\node (B) [right of=A]  {$B$};
		\node (C) [right of=B]  {$C$};
		\path (B) edge  node {} (C);		
		\draw[semithick,-] (A) -- ++(.25cm,0) |- (.4,0);
		\draw [semithick,->](x) edge  node {} (B);
		\end{tikzpicture}
	}\quad
	\subfigure[Sync2Drain]{
		\centering \begin{tikzpicture}[->,>=stealth,font=\sffamily,semithick,node distance=1.8cm]
		\node (A)  {$A$};
		\node (B) [right of=A]  {$B$};
		\node (C) [right of=B]  {$C$};
		\path (B) edge  node {} (A)
		(B) edge [->,>=to reversed,font=\sffamily,semithick] node {} (C)
		(C) edge [->,>=to reversed,font=\sffamily,semithick] node {} (B);
		
		\end{tikzpicture}
	}
	\caption{Examples of Reo models}
\label{fig:reoModels}
\end{figure*}
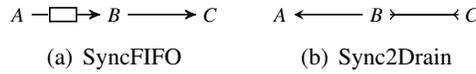

The next definition maps each canonical connector that composes a Reo model to a \relo\ program. The left hand side of each mapping rule in Definition~\ref{def:parseBaseCases} is the atomic Reo connector, while the right hand size is the resulting \relo\ atomic program $\pi_i = (f_i,b_i)$, with the same behaviour as of the Reo connector. 

\begin{definition}[$parse$ base cases] \label{def:parseBaseCases} Each canonical Reo connector is mapped to a \relo\ program in $parse$:
\begin{itemize}[noitemsep]
	\item \scalebox{.8}{$\begin{tikzpicture}[->,>=stealth',font=\sffamily,semithick,node distance=1.6cm]
	\node (A) {$A$};
	\node (B) [right of=A]  {$B$};
	\path (A) edge  node {} (B);	
	\end{tikzpicture}$} to $A \to B$
	
	\item \scalebox{.8}{$\begin{tikzpicture}[->,>=stealth',font=\sffamily,semithick,node distance=1.6cm,dashed]
	\node (A)  {$A$};
	\node (B) [right of=A]  {$B$};
	\path (A) edge  node {} (B);
	\end{tikzpicture}$} to $(A,A \to B)$

	\item \scalebox{.8}{$\begin{tikzpicture}[->,>=stealth,font=\sffamily,semithick,node distance=1cm]
		\node (A)  {$A$};
		\node (x) [draw,rectangle,minimum width=0.5cm,minimum height=0.2cm] at  (.76,0) {};
		\node (B) [right of=x]  {$B$};
		\draw[semithick,-] (A) -- ++(.6cm,0) |- (.4,0);
		\draw [semithick,->](x) edge  node {} (B);
		\end{tikzpicture}$} to $fifo(A,B)$
	
	\item \scalebox{.8}{$\begin{tikzpicture}[->,>=to reversed,font=\sffamily,semithick,node distance=1.8cm]
		\node (A)  {$A$};
		\node (B) [right of=A]  {$B$};
		\path (A) edge  node {} (B)
		(B) edge  node {} (A);
		\end{tikzpicture}$} to $SBlock(A,B)$
	
	\item \scalebox{.8}{$\begin{tikzpicture}[->,>=to reversed,font=\sffamily,semithick,node distance=1.8cm]
		\node (A)  {$A$};
		\node (B) [right of=A]  {$B$};
		\path (A) edge  node {\tikz \draw[|-|,semithick ] (0,0) -- +(.1,0);} (B)
		(B) edge  node {} (A);
		\end{tikzpicture}$} to $ABlock(A,B)$
	
	\item \scalebox{.8}{$\begin{tikzpicture}[->,>=stealth,font=\sffamily,semithick,node distance=1.8cm,] 
		\node (A)  {$A$};
		\node (B) [right of=A]  {$B$};
		\path (A) edge node {\tikz \draw[-triangle 90] (0,0) -- +(.1,0);} (B);
		\end{tikzpicture}$} to $Transform(f,A,B)$, $f \colon Data \to Data$ is a transformation function.
	
	\item \scalebox{.8}{$\begin{tikzpicture}[->,decorate,decoration=snake,>=stealth,font=\sffamily,semithick,node distance=1.8cm]
		\node (A)  {$A$};
		\node (B) [right of=A]  {$B$};
		\draw [->,decorate,decoration=snake] (A) -- (B);
		\end{tikzpicture}$} to $Filter(P,A,B)$, $P$ is a logical predicate over the data item in $A$.
	
	\item \scalebox{.8}{$\begin{tikzpicture}[>=stealth,font=\sffamily,semithick,node distance=.8cm]
		\node (A) at (0,0) {$A$};
		\node (B) at (0,0.8)  {$B$};
		\node (x) [fill,circle,inner sep=1pt] at (0.8,0.4) {};
		\node (C) [right of=x] {$C$};
		\path (A) edge  node {} (x)
		(B) edge  node {} (x)
		(x) edge  node {} (C); 
		\end{tikzpicture}$} to $(A \to C, B \to C)$
	
	\item \scalebox{.8}{$\begin{tikzpicture}[>=stealth,font=\sffamily,semithick,node distance=1cm]
		\node (A) at (0,0) {$A$} ;
		\node (x) [fill,circle,inner sep=1pt] at (0.8,0) {}; 
		\node (B) at (1.6,-0.4) {$B$};
		\node (C) at (1.6,0.4)  {$C$};
		\path (A) edge  node {} (x)
		(x) edge  node {} (B)
		(x) edge node  {} (C);
		\end{tikzpicture}$} to $(A \to B, A \to C)$
\end{itemize}
\end{definition}

Considering that each \relo\ program $\Pi$ is the composition of programs $\pi_1 \odot \pi_2, \odot \dots \odot \pi_n, \pi_i = (f_i,b_i)$ as Reo programs, $parse$ is formalized in Definition~\ref{def:parse}. The symbol $\circ$ denote the addition of an element to $s$, the resulting set of $parse$'s processing.

\begin{definition}[$parse$ function] \label{def:parse} The function that interprets the execution of a \relo\ program is defined as $parse(f, b, s)$. We define $\varepsilon$ as an abbreviation to denote when there is no \relo\ program left to process (i.e. the base case when no program is parametrized). Its outcome is detailed as below.

\begin{itemize}[noitemsep]
	\item $s, \text{ if } f = b = \varepsilon$
	\item $ 
	parse(f_j,b,s \circ A \to B), \text{ if } f = \begin{tikzpicture}[->,>=stealth',font=\sffamily,semithick,node distance=1.6cm,baseline=-0.85ex]
	\node (A) {$A$};
	\node (B) [right of=A]  {$B$};
	\path (A) edge  node {} (B);	
	\end{tikzpicture} \odot f_j
	$\begin{itemize}
		\item$s \circ A \to B, \text{ if } f = \begin{tikzpicture}[->,>=stealth',font=\sffamily,semithick,node distance=1.6cm]
		\node (A) {$A$};
		\node (B) [right of=A]  {$B$};
		\path (A) edge  node {} (B);	
		\end{tikzpicture}$
	\end{itemize}
	\item$ 
	parse(f_j,b,s \circ (A,A \to B)), \text{ if } f = \begin{tikzpicture}[->,>=stealth',font=\sffamily,semithick,node distance=1.6cm,dashed]
	\node (A)  {$A$};
	\node (B) [right of=A]  {$B$};
	\path (A) edge  node {} (B);
	\end{tikzpicture} \odot f_j
	$\begin{itemize}
		\item $s \circ (A,A \to B), \text{ if } f = \begin{tikzpicture}[->,>=stealth',font=\sffamily,semithick,node distance=1.6cm,dashed]
		\node (A)  {$A$};
		\node (B) [right of=A]  {$B$};
		\path (A) edge  node {} (B);
		\end{tikzpicture}$
	\end{itemize}
	\item $ 
	parse(f_j,b,s) \circ fifo(A,B), \text{ if } f = \begin{tikzpicture}[->,>=stealth,font=\sffamily,semithick,node distance=1cm]
	\node (A)  {$A$};
	\node (x) [draw,rectangle,minimum width=0.5cm,minimum height=0.2cm] at  (.95,0) {};
	\node (B) [right of=x]  {$B$};
	\draw[semithick,-] (A) -- ++(.7cm,0) |- (.5,0);
	\draw [semithick,->](x) edge  node {} (B);
	\end{tikzpicture} \odot f_j
	$\begin{itemize}
	\item $ (s \circ fifo(A,B)), \text{ if } f = \begin{tikzpicture}[->,>=stealth,font=\sffamily,semithick,node distance=1cm]
		\node (A)  {$A$};
		\node (x) [draw,rectangle,minimum width=0.5cm,minimum height=0.2cm] at  (.95,0) {};
		\node (B) [right of=x]  {$B$};
		\draw[semithick,-] (A) -- ++(.7cm,0) |- (.5,0);
		\draw [semithick,->](x) edge  node {} (B);
		\end{tikzpicture}$
	\end{itemize}
	\item $ 
	SBlock(A,B) \circ parse(f, b_j,s), \text{ if } b = \begin{tikzpicture}[->,>=to reversed,font=\sffamily,semithick,node distance=1.8cm]
	\node (A)  {$A$};
	\node (B) [right of=A]  {$B$};
	\path (A) edge  node {} (B)
	(B) edge  node {} (A);
	\end{tikzpicture} \odot b_j
	$\begin{itemize}
		\item $ (SBlock(A,B) \circ s), \text{ if } b = \begin{tikzpicture}[->,>=to reversed,font=\sffamily,semithick,node distance=1.8cm]
		\node (A)  {$A$};
		\node (B) [right of=A]  {$B$};
		\path (A) edge  node {} (B)
		(B) edge  node {} (A);
		\end{tikzpicture}$
	\end{itemize}
	\item $
	ABlock(A,B) \circ parse(f, b_j,s), \text{ if } b = \begin{tikzpicture}[->,>=to reversed,font=\sffamily,semithick,node distance=1.8cm]
	\node (A)  {$A$};
	\node (B) [right of=A]  {$B$};
	\path (A) edge  node {\tikz \draw[|-|,semithick ] (0,0) -- +(.1,0);} (B)
	(B) edge  node {} (A);
	\end{tikzpicture} \odot b_j $
	\begin{itemize}
	\item $ (ABlock(A,B) \circ s),\text{ if } b = \begin{tikzpicture}[->,>=to reversed,font=\sffamily,semithick,node distance=1.8cm]
	\node (A)  {$A$};
	\node (B) [right of=A]  {$B$};
	\path (A) edge  node {\tikz \draw[|-|,semithick ] (0,0) -- +(.1,0);} (B)
	(B) edge  node {} (A);
	\end{tikzpicture}$
	\end{itemize}
	\item $ 
	parse(f_j,b,s \circ Transform(f,A,B)), \text{ if } f =  \begin{tikzpicture}[->,>=stealth,font=\sffamily,semithick,node distance=1.8cm,] 
	\node (A)  {$A$};
	\node (B) [right of=A]  {$B$};
	\path (A) edge node {\tikz \draw[-triangle 90] (0,0) -- +(.1,0);} (B);
	\end{tikzpicture} \odot f_j$
	\begin{itemize}
	\item $ (Transform(f,A,B) \circ s), \text{ if } f =  \begin{tikzpicture}[->,>=stealth,font=\sffamily,semithick,node distance=1.8cm,] 
	\node (A)  {$A$};
	\node (B) [right of=A]  {$B$};
	\path (A) edge node {\tikz \draw[-triangle 90] (0,0) -- +(.1,0);} (B);
	\end{tikzpicture}$ 
	\end{itemize}
	\item $ 
	parse(f_j,b,s \circ Filter(P,A,B)), \text{ if } f = \begin{tikzpicture}[->,decorate,decoration=snake,>=stealth,font=\sffamily,semithick,node distance=1.8cm]
	\node (A)  {$A$};
	\node (B) [right of=A]  {$B$};
	\draw [->,decorate,decoration=snake] (A) -- (B);
	\end{tikzpicture} \odot f_j $
	\begin{itemize}
	\item $ (Filter(P,A,B) \circ s), \text{ if } f = \begin{tikzpicture}[->,decorate,decoration=snake,>=stealth,font=\sffamily,semithick,node distance=1.8cm]
	\node (A)  {$A$};
	\node (B) [right of=A]  {$B$};
	\draw [->,decorate,decoration=snake] (A) -- (B);
	\end{tikzpicture}$
	\end{itemize}
	\item $ 
	parse(f_j,b,s\circ (A \to C, B \to C)), \text{ if } f = \begin{tikzpicture}[>=stealth,font=\sffamily,semithick,node distance=.8cm, baseline=0.5mm]
	\node (A) at (0,0) {$A$};
	\node (B) at (0,0.8)  {$B$};
	\node (x) [fill,circle,inner sep=1pt] at (0.8,0.4) {};
	\node (C) [right of=x] {$C$};
	\path (A) edge  node {} (x)
	(B) edge  node {} (x)
	(x) edge  node {} (C); 
	\end{tikzpicture} \odot f_j$
	\begin{itemize}
	\item $ (s \circ (A \to C, B \to C)), \text{ if } f = \begin{tikzpicture}[>=stealth,font=\sffamily,semithick,node distance=.8cm, baseline=0.5mm]
	\node (A) at (0,0) {$A$};
	\node (B) at (0,0.8)  {$B$};
	\node (x) [fill,circle,inner sep=1pt] at (0.8,0.4) {};
	\node (C) [right of=x] {$C$};
	\path (A) edge  node {} (x)
	(B) edge  node {} (x)
	(x) edge  node {} (C); 
	\end{tikzpicture}$
	\end{itemize}
	\item 
	$parse(f_j,b,s\circ (A \to B, A \to C)), \text{ if } f = \begin{tikzpicture}[>=stealth,font=\sffamily,semithick,node distance=1cm, baseline=0.3mm]
	\node (A) at (0,0) {$A$} ;
	\node (x) [fill,circle,inner sep=1pt] at (0.8,0) {}; 
	\node (B) at (1.6,-0.4) {$B$};
	\node (C) at (1.6,0.4)  {$C$};
	\path (A) edge  node {} (x)
	(x) edge  node {} (B)
	(x) edge node  {} (C);
	\end{tikzpicture} \odot f_j$
	\begin{itemize}
	\item $ (s \circ (a \to b, a \to c)), \text{ if } f = \begin{tikzpicture}[>=stealth,font=\sffamily,semithick,node distance=1cm, baseline=0.3mm]
	\node (A) at (0,0) {$A$} ;
	\node (x) [fill,circle,inner sep=1pt] at (0.8,0) {}; 
	\node (B) at (1.6,-0.4) {$B$};
	\node (C) at (1.6,0.4)  {$C$};
	\path (A) edge  node {} (x)
	(x) edge  node {} (B)
	(x) edge node  {} (C); \end{tikzpicture}$
	\end{itemize}
\end{itemize}
\end{definition}

We employ $parse$ to interpret Reo programs $\Pi$ as a sequence of occurrences of possible data flow (where each flow corresponds to the execution of a Reo connector). These data flow may denote data transfer (\relo\ programs ($A \to B$) and ($A$,$A \to B$), flow ``blocks'' induced by connectors such as SyncDrain and aSyncDrain (\relo\ programs $SBlock(A,B)$ and $ABlock(A,B)$ --- the first one requires that data flow synchronously through its ports, while the latter requires that data flow asynchronously through its ports). There is also the notion of a buffer introduced by FIFO connectors (\relo\ program $fifo(A,B)$), which data flow into a buffer before flowing out of the channel, and merging/replicating data flow between ports, respectively denoted by channels Merger and Replicator (\relo\ programs $(A \to C, B \to C)$ and $(A \to B, A \to C)$ respectively). 

There are also special data flows, denoting the ``transformation'' of some data flowing from A to B as $Transform(f,A,B)$ which will apply $f$ with the data in $A$ before it sends $f(D_A)$ ($D_A$ denoting the data item in A) to $B$, and the filtering of data flow by some predicate as $Filter(P,A,B)$, $P$ as a quantifiable-free predicate over the data item seen in $A$. Therefore, data will flow to $B$ only if $P(D_A)$ is satisfied.

After processing $\pi$ with $parse$, the interpretation of the execution of $\pi$ is given by $go(t,s,acc), go\colon s \times s \to s$, where $s$ is a string denoting the processed program $\pi$ as the one returned by $parse$, and $t$ is the initial data flow of ports of the Reo program $\pi$. The parameter $acc$ holds all connectors of the Reo circuit that satisfy their respective required conditions for data to flow. In what follows we define $ax \prec t$ as an operator which states that $ax$ is in $t$, $ax$ a single data of a port and $t$ a structure containing data flows for ports $p \in \mathcal{N}$.

Example~\ref{def:gExample} shows how $parse$ functions and illustrates why it is necessary. The programs that depict the FIFO connectors from Fig.~\ref{fig:SequencerReo} are the last programs to be executed, while the ones that denote ``immediate'' flow (the Sync channels) come first. This is done to preserve the data when these connectors fire (if eligible). Suppose that there is a data item in the buffer between X and Y and a data item in Y (i.e., $t = X1Y, Y0$). If the data item leaves the buffer first then the data item in Y, the latter will be overwritten and the information is lost.

\begin{example} \label{def:gExample}
	let $\pi$ be the Reo program corresponding to the circuit in Fig.~\ref{fig:SequencerReo}:\\
	$\pi$ = \scalebox{0.7}[0.7]{ $\rfifo{X}{Y} \odot \rsync{Y}{A} \odot \rfifo{Y}{W} \odot \rsync{W}{B} \odot \rfifo{W}{Z} $}\\ \scalebox{0.7}[0.7]{ $\odot \rsync{Z}{C} \odot \rsync{Z}{X}$}\\
	$parse(\pi,\{\})$ = \{$Y \to A; W \to B ; Z \to C ; Z \to X ; fifo(X,Y); fifo(Y,W);fifo(W,Z)$\}
\end{example}

The usage of $parse$ is required to eliminate problems regarding the execution order of $\pi$'s Reo channels, which could be caused by processing $\pi$ the way it is inputted (i.e., its connectors can be in any order). Consider, for example, the behavior of SyncDrain and aSyncDrain programs as ``blocking'' programs as discussed earlier. In a single step, they must be evaluated before the flow programs, because if they fail to execute due to missing requirements, data should not flow from their port names to other connectors. In a nutshell, $parse$ organizes the program so this verification can be performed.

Therefore, the interpretation of a $\pi$ program processed by $parse$ is performed by $go(t,s,acc)$, where $s$ is a string containing $\pi$ as processed by $parse$, $t$ is $\pi$'s initial data flow, and $acc$ filters the connectors of the \relo\ program that can be fired if the requirements to do so are met.

Definition \ref{def:go} will check for each of the Reo connectors processed by $parse$ satisfies the required condition to fire, following the connectors' behaviour. Operator $\prec$ denotes whether the data flow is within the current data flow $t$ being evaluated. It is also used to denote whether the program currently being evaluated in $s$ repeats in $\Pi$. Operator $\setminus$ denotes the removal of an connector from the accumulator $acc$.

\begin{definition}[Relation $go$ for a single execution step] \label{def:go} We define $go(t,s,acc)$ as follows:
	
	\begin{itemize}[noitemsep]
		
		\item $s = \epsilon : fire(t,acc)$
		
		\item $s = A \to B \circ s' : $\begin{itemize}[noitemsep]
			\item $go(t,s',acc \circ (A \to B)), \text{ iff }  Ax \prec t, (A \to B) \nprec s'$ 
			\item $go(t,s', (acc \circ (A \to B))\setminus s'_{j}) \cup  go(t, s',acc), \text{ iff } \begin{cases}
			Ax \prec t,\\
			(A \to B) \nprec s' \\
			\exists s'_{j} \in acc \mid sink(s'_{j}) = B \end{cases}$
			\item $go(t,s',acc)$, otherwise
		\end{itemize}
		
		\item $s = (A,A \to B) \circ s' : $\begin{itemize}[noitemsep]
			\item $go(t,s',acc \circ (A \to B)) \cup go(t,s',acc \circ (A \to A)) , \text{ iff }  Ax \prec t, (A \to B) \nprec s'$ 
			\item $go(t,s', (acc \circ (A \to B))\setminus s'_{j}) \cup  go(t, s',acc), \text{ iff } \begin{cases}
			Ax \prec t,\\
			(A \to B) \nprec s' \\
			\exists s'_{j} \in acc \mid sink(s'_{j}) = B \end{cases}$
			\item $go(t,s',acc)$, otherwise
		\end{itemize}
		
		\item $s = fifo(A,B) \circ s' :$
		\begin{itemize}[noitemsep]
			
			\item $go(t,s',acc \circ (AxB)), \text{ iff } Ax \prec t, fifo(A,B) \nprec s',(AxB) \nprec acc$
			
			\item $go(t,s', acc \circ (AxB \to Bx)), \text{ iff } AxB \prec t, fifo(A,B) \nprec s'$
			
			\item $go(t,s',(acc \circ (AxB \to Bx) )\setminus s'_{j}) \cup  go(t, s',acc), \text{ iff } \begin{cases} AxB \prec t,\\ fifo(A,B) \nprec s',\\ \exists s'_{j} \in acc \mid sink(s'_{j}) = B\\\end{cases}$ 
			\item $go(t,s',acc)$, otherwise
		\end{itemize}
		
		\item 	$s = Sblock(A,B) \circ s' :$ \begin{itemize}[noitemsep]
			\item $go(t,s',acc), \text{ iff } \begin{cases}  (Ax \prec t \land Bx \prec t) \lor (Ax \nprec t \land Bx \nprec t) \\ Sblock(A,B) \nprec s' \end{cases}$ 
			\item $go(t,halt(A,B,s'),acc), \text{ iff } \begin{cases} (Ax \prec t \land Bx \nprec t) \lor (Ax \nprec t \land Bx \prec t)\\  Sblock(A,B) \nprec s' \end{cases}$
		\end{itemize}
		
		\item $s = Ablock(A,b)\circ s':$
		\begin{itemize}[noitemsep]
			\item $go(t,s',acc), \text{ iff } \begin{cases} (Ax \nprec t \land Bx \prec t) \lor (Ax \prec t \land Bx \nprec t) \lor \\ (Ax \nprec t \land Bx \nprec t), Ablock(A,B) \nprec s'\end{cases}$ 
			\item $go(t,halt(A,B,s'),acc), \text{ iff } \begin{cases} (Ax \prec t \land Bx \prec t), \\  Ablock(A,B) \nprec s' \end{cases}$
		\end{itemize}
		
		\item $s = Transform(f,A,B) \circ s' : $\begin{itemize}[noitemsep]
			\item $go(t,s',acc \circ (f(D_A) \to B)), \text{ iff }\begin{cases} 
			ax \prec t\\ 
			Transform(f,A,B) \nprec s' \end{cases}$ 
			\item $go(t,s', (acc \circ (f(D_A) \to B))\setminus s'_{j}) \cup  go(t, s',acc), \text{ iff } \begin{cases}
			Ax \prec t,\\
			Transform(f,A,B) \nprec s' \\
			\exists s'_{j} \in acc \mid sink(s'_{j}) = B \end{cases}$
			\item $go(t,s',acc)$, otherwise
		\end{itemize}
		
		\item $s = Filter(f,A,B) \circ s' : $\begin{itemize}[noitemsep]
			\item $go(t,s',acc \circ (A \to B)), \text{ iff }\begin{cases} 
			Ax \prec t\\ 
			P(D_A) \text{ holds }\\
			Filter(f,A,B) \nprec s' \end{cases}$ 
			\item $go(t,s', (acc \circ (A \to B))\setminus s'_{j}) \cup  go(t, s',acc), \text{ iff } \begin{cases}
			Ax \prec t,\\
			P(D_A) \text{ holds }\\
			Filter(f,A,B) \nprec s' \\
			\exists s'_{j} \in acc \mid sink(s'_{j}) = B \end{cases}$
			\item $go(t,s',acc)$, otherwise
		\end{itemize}
		

		%
		%
		
	\end{itemize}
\end{definition}

The existing condition after each return condition of $go$ denotes the case where two or more Reo connectors within a circuit have the same sink node. This implies that if both of their respective source nodes have data flowing simultaneously, their sink nodes will have data flowing nondeterministically. Such condition models this scenario, considering when both cases may happen as two nondeterministic ``distinct'' possible executions. Therefore, the operation $acc \circ (X \to Y))\setminus s'_{j}$ removes every interpretation of $s'$ which sink node equals $Y$, while $go(t, s',acc)$ denotes an execution containing the removed $s'_{j}$ but not considering $X \to Y$. The return condition $s = \epsilon$ denotes that the program as a whole has already been processed.

Considering the cases including block programs induced by SyncDrain and AsyncDrain connectors, $halt(A,B,s')$ is defined as a supporting function that will be used in the case the block program conditions fail. Then, data flow that was in the ports of the SyncDrain/AsyncDrain evaluated cannot be further considered in this execution steps: channels that have their sink node pointed to $A$ or $B$.

Intuitively, $go$ is a function that processes a program $\pi$ with input $t$ as the program's data initially available at ports $p \in \pi$ and returns the next data configuration after processing all connectors and verifying whether they are eligible for data to flow. The return of $go$ depends on a function $fire$ which is bound to return the final configuration of the Reo circuit after an iteration (i.e., the last ports that data flow). We define $sink(s'_j)$ as the sink node of a connector, in this case, the port name where a data item flowing into a Reo connector is bound to. The operation denoted by $\cup$ is the standard set union.

Definition $go$ employs a function named $fire\colon T \times s \rightarrow T$ which returns the firing of all possible data flows in the Reo connector, given the Reo program $\pi$ and an initial data flow on ports of $\pi$. The set $T$ is the set of possible data flows as constructed by the BNF grammar in Definition~\ref{def:reloLanguage}. The function $fire$ returns the resulting data flow of this execution step by considering the program processed by $go$ as $s$ and the current step's data flow $t$. Parameter $s$ contains \relo\ programs as yielded by $parse$.

\begin{definition}[Data marking relation $fire$] \label{def:fire}


\begin{align}
fire(t,s) = \begin{cases}
\epsilon, \text{ if } s = \epsilon \\
AxB \circ fire(t,s'), \text{ if } s = (AxB) \circ s' \text{ and } Ax \prec t\\
B(f(a)) \circ fire(t,s'), \text{ if } s = (f(D_A) \to B) \circ s' \text{ and } Ax \prec t\\
Bx \circ fire(t,s'), \text{ if } \begin{cases} 
s = (A \to B) \circ s' \text{ and } Ax \prec t, or\\
s = (AxB \to Bx) \circ s' \text{ and } axb \prec t\\
\end{cases}\\
\end{cases}
\end{align}

\end{definition}

We define $f_{ReLo}$ as the transition relation of a \relo\ model. It denotes how the transitions of the model fire, i.e., given an input $t$ and a program $\pi$ denoting a Reo circuit, $f_{ReLo}(t,\pi)$ interfaces with $go$ to return the resulting data flow of $\pi$ given that data depicted by $t$ are flowing in the connector's ports.

\begin{definition}{Transition relation} \label{def:f}
$f_{ReLo}(t,\pi) = go(t,(parse(\pi,[])),[])$

\end{definition}

We define $f_{ReLo}(t,\pi^\star)$ as the application of $f_{ReLo}(t,\pi)$ iteratively for the (nondeterministic finite) number of steps denoted by $\star$, starting with $t$ with $\pi$, and considering the obtained intermediate $t'$ in the steps.

A \relo\ frame is a structure based on Kripke frames~\cite{Kripke1959} formally defined as a tuple $\mathcal{F} = \langle S, \Pi ,R_\Pi, \delta,$ $\lambda \rangle$, where each element of $\mathcal{F}$ is described by Definition~\ref{def:reloFrame}.

\begin{definition}[\relo\ frame] \label{def:reloFrame}  $S$ is a non-empty enumerable set of states and $\Pi$ a Reo program.
	\begin{itemize}[noitemsep]
		\item $R_\Pi \subseteq S \times S$ is a relation defined as follows.
		\begin{itemize}[noitemsep]
			\item $R_{\pi_i} =	\lbrace uR_{\pi_i}v \mid f_{ReLo}(t,\pi_i) \prec \delta(v)$, $t \prec \delta(u) \rbrace$, $\pi_i$ is any combination of any atomic program which is a subprogram of $\Pi$.
			\item $R_{\pi_i^\star} = R^\star_{\pi_i}$, the reflexive transitive closure (RTC) of $R_{\pi_i}$.	
		\end{itemize}
		\item $\lambda \colon S \times \mathcal{N} \rightarrow \mathbbm{R}$ is a function that returns the time instant a data item in a data markup flows through a port name of $\mathcal{N}$.
		\item $\delta \colon S \rightarrow T$, is a function that returns data in ports of the circuit in a state $s \in S$, $T$ being the set of possible data flows in the model.

	\end{itemize}
\end{definition}

From Definition~\ref{def:reloFrame}, a \relo\ model is formally defined as a tuple $\mathcal{M} = \langle \mathcal F,  \textbf{V} \rangle$ by Definition~\ref{def:reloModels}. Intuitively, it is a tuple consisting of a \relo\ frame and a valuation function, which given a state $w$ of the model and a propositional symbol $\varphi \in \Phi$, maps to either $true$ or $false$.

\begin{definition}[\relo\ models] \label{def:reloModels} A model in \relo\ is a tuple $\mathcal{M} = \langle \mathcal F,  \textbf{V} \rangle$, where $\mathcal{F}$ is a \relo\ frame and $V\colon S \times \Phi \to \{true,false\}$ is the model's valuation function
\end{definition}

%


\begin{definition}[Satisfaction notion] \mbox{}
\begin{itemize}[noitemsep]
	\item $\mathcal{M}{,}s \Vdash p \text{ iff } V{(s,p)} = true$
	\item $\mathcal{M}{,}s \Vdash \top$ always
	\item $\mathcal{M}{,}s \Vdash \neg \varphi \text{ iff } \mathcal{M},s \nVdash \varphi$
	\item $\mathcal{M}{,}s \Vdash \varphi_1 \land \varphi_2 \text{ iff } \mathcal{M}{,}s \Vdash \varphi_1$ and $\mathcal{M}{,}s \Vdash \varphi_2$
	\item $\mathcal{M}{,}s \Vdash \langle t, \pi \rangle \varphi$ if there exists a state $w \in S$, $sR_{\pi}w$, and $\mathcal{M}{,}s \Vdash \varphi$
\end{itemize}
We denote by $\mathcal M\Vdash\varphi$ if $\varphi$ is satisfied in all states of $\mathcal M$. By $\Vdash\varphi$ we denote that $\varphi$ is valid in any state of any model.
\end{definition}

We recover the circuit in Fig.~\ref{fig:SequencerReo} as an example. Let us consider s = $D_X$, (i.e. t = D1) and the Sequencer's corresponding model $\mathcal{M}$. Therefore, $\mathcal{M},D_X \Vdash \langle t, \pi \rangle p$ holds if $V(D_{XfifoY}, p) = true$ as $D_{XfifoY}$ is the only state where $D_X R_\Pi D_{XfifoY}$. For example, one might state $p$ as ``There is no port with any data flow'', hence $V(D_{XfifoY}, p) = true$.

As another usage example, we formalize some properties which may be interesting for this connector to have. Let us consider that the data markup is $t = X1$, $\mathcal{M}$ the model regarding the Sequencer, and the states' subscript denoting which part of the connector has data. The following example state that for this data flow, after every single execution of $\pi$, it is not the case that the three connected entities have their data equal to $1$ simultaneously, but it does have data in its buffer from $X$ to $Y$. 


\begin{example}
	
	\noindent$\lbrack X1,\pi \rbrack \neg (D_A = 1 \land D_B = 1 \land D_C = 1) \land t' = X1Y$, where $t' = f_{ReLo}(t,\pi)$\\
	$\mathcal{M}{,}D_X \Vdash \lbrack X1,\pi \rbrack \neg (D_A = 1 \land D_B = 1 \land D_C = 1) \land t' = X1Y$.\\
	$\mathcal{M}{,}D_{\scalebox{0.5}[0.5]{\begin{tikzpicture}[->,>=stealth,font=\sffamily,semithick,node distance=1cm]
			\node (A)  {$X$};
			\node (x) [draw,rectangle,minimum width=0.5cm,minimum height=0.2cm] at  (.95,0) {}; 
			\node (B) [right of=x]  {$Y$};
			\draw[semithick,-] (A) -- ++(.7cm,0) |- (.5,0);
			\draw [semithick,->](x) edge  node {} (B);
			
\end{tikzpicture}}} \Vdash  \neg (D_A = 1 \land D_B = 1 \land D_C = 1) \land t' = X1Y$.\\
$\mathcal{M}{,}D_{\scalebox{0.5}[0.5]{\begin{tikzpicture}[->,>=stealth,font=\sffamily,semithick,node distance=1cm]
		\node (A)  {$X$};
		\node (x) [draw,rectangle,minimum width=0.5cm,minimum height=0.2cm] at  (.95,0) {}; 
		\node (B) [right of=x]  {$Y$};
		\draw[semithick,-] (A) -- ++(.7cm,0) |- (.5,0);
		\draw [semithick,->](x) edge  node {} (B);
		
		\end{tikzpicture}}}
\Vdash  \neg (D_A = 1 \land D_B = 1 \land D_C = 1) \text{ and } 
\mathcal{M},D_{\scalebox{0.5}[0.5]{\begin{tikzpicture}[->,>=stealth,font=\sffamily,semithick,node distance=1cm]
		\node (A)  {$X$};
		\node (x) [draw,rectangle,minimum width=0.5cm,minimum height=0.2cm] at  (.95,0) {}; 
		\node (B) [right of=x]  {$Y$};
		\draw[semithick,-] (A) -- ++(.7cm,0) |- (.5,0);
		\draw [semithick,->](x) edge  node {} (B);
		
		\end{tikzpicture}}} \Vdash t' = X1Y$.\\

\end{example}

The notion of $\mathcal{M},D_X \Vdash \langle t, \pi^\star \rangle p$ holds if a state $s$ is reached from $D_X$ by means of $R_\pi^\star$ with $V(s,p) = \top$. If we state $p$ as ``the data item of port $X$ equals $1$'', it holds because $D_X R_\pi^\star D_X$ and $V(D_X, p) = \top$. If there is an execution of $\pi$ that lasts a nondeterministic finite number of iterations, and there is data in $C$ equals to $1$, then there is an execution under the same circumstances where the same data has been in $B$.

\begin{example}
$\langle t,\pi^\star \rangle D_C = 1 \rightarrow \langle t, \pi^\star \rangle D_B = 1 $ \\
$\mathcal{M}{,}D_X \Vdash \langle t,\pi^\star \rangle D_C = 1 \rightarrow \langle t, \pi^\star \rangle D_B = 1 $\\
$\mathcal{M}{,}D_X \Vdash \neg( \langle t,\pi^\star \rangle D_C = 1) \lor \langle t, \pi^\star \rangle D_B = 1 $\\
$\mathcal{M}{,}D_X \Vdash \lbrack t,\pi^\star \rbrack \neg D_C = 1 \lor \langle t, \pi^\star \rangle D_B = 1 $\\
$\mathcal{M}{,}D_X \Vdash \lbrack t,\pi^\star \rbrack
\neg D_C = 1  \text{ or } \mathcal{M}{,}D_X \Vdash \langle t, \pi^\star \rangle D_B = 1 $\\
$\mathcal{M}{,}D_X \Vdash \langle t, \pi^\star \rangle D_B = 1 $, because 
$\mathcal{M}{,}D_B \Vdash  D_B = 1 $ and $D_X R_{\pi^\star} R_B$. \\

\end{example}

\subsection{Axiomatic System} \label{subsec:axioms}

We define an axiomatization of \relo, discuss its soundness and completeness. 

\begin{definition}[Axiomatic System] \label{def:axiomaticSystem} \mbox{}\\
	
	\begin{varwidth}[T]{.5\textwidth}
		\begin{itemize}[noitemsep]
			\item[\textbf{(PL)}] Enough Propositional Logic tautologies 
			
			\item[\textbf{(K)}] $\lbrack t , \pi \rbrack (\varphi \rightarrow \psi ) \rightarrow (\lbrack t, \pi \rbrack \varphi \rightarrow \lbrack  t, \pi \rbrack \psi)$ 
			
			\item[\textbf{(And)}] $\lbrack t, \pi \rbrack (\varphi \land \psi) \leftrightarrow \lbrack t, \pi \rbrack \varphi \land \lbrack t, \pi \rbrack \varphi$
			
			\item[\textbf{(Du)}] $\lbrack  t, \pi \rbrack \varphi \leftrightarrow \neg\langle t, \pi \rangle \neg \varphi$
			
			\item[\textbf{(R)}] $\langle t, \pi \rangle \varphi \leftrightarrow \varphi$ iff $f_{ReLo}(t,\pi) = \epsilon$
			
			\item[\textbf{(It)}] $\varphi \land \lbrack  t, \pi \rbrack \lbrack  t_{(f,b)}, \pi^\star \rbrack \varphi \leftrightarrow \lbrack  t, \pi^\star \rbrack \varphi$, $t_{(f,b)} = f_{ReLo}(t,\pi)$
			
			\item[\textbf{(Ind)}] $\varphi \land \lbrack  t, \pi^\star \rbrack(\varphi \rightarrow \lbrack  t_{(f,b)^\star}, \pi \rbrack \varphi) \rightarrow \lbrack  t, \pi^\star \rbrack \varphi$, $t_{(f,b)^\star} = f_{ReLo}(t,\pi^\star)$

		\end{itemize}	
	\end{varwidth}
	\hspace{4em}
	\begin{varwidth}[T]{.5\textwidth}
		\begin{itemize}[noitemsep]

			\item[\textbf{(MP)}] \begin{prooftree}
				\AxiomC{$\varphi$} \AxiomC{$\varphi \rightarrow \psi$}
				\BinaryInfC{$\psi$} 
			\end{prooftree}
			
			\item[\textbf{(Gen)}] \begin{prooftree}
				\AxiomC{$\varphi$}
				\UnaryInfC{$\lbrack t,\pi \rbrack \varphi$}
			\end{prooftree}

		\end{itemize}
	\end{varwidth}
\end{definition}
\begin{lemma}[Soundness] \label{lemma:soundnessRelo}
		\begin{proof} \mbox{}\\ Axioms \textbf{(PL)}, \textbf{(K)}, \textbf{(And)} and \textbf{(Du)} are standard in Modal Logic literature, along with rules \textbf{(MP)} and \textbf{(Gen)}~\cite{Harel2001}. Axiom \textbf{(It)} and \textbf{(Ind)} are similar from PDL.
			\textbf{(R):} $\langle t, \pi \rangle \varphi \leftrightarrow \varphi$ iff $f_{ReLo}(t,\pi) = \epsilon$\\
			Suppose by contradiction that exists a state $s$ from a model $\mathcal M = \langle S, \Pi, R_\Pi, \delta, \lambda, V \rangle$ where \textbf{(R)} does not hold. There are two possible cases.\\
			($\Rightarrow$)
			Suppose by contradiction $\mathcal{M},s \Vdash \langle t, (f,b) \rangle \varphi $ and $\mathcal{M},s \nVdash \varphi$. $\mathcal{M},s \Vdash \langle t, (f,b) \rangle \varphi $ iff there is a state $v \in S$ such that $s R_{\pi} v$. Because $f_{ReLo}(t,(f,b)) = \epsilon, s = v$ (i.e., in this execution no other state is reached from $s$). Therefore,  $\mathcal{M},s \Vdash \varphi$, contradicting $\mathcal{M},s \nVdash \varphi$.\\
			($\Leftarrow$)
			Suppose by contradiction $\mathcal{M},s \Vdash \varphi$ and $\mathcal{M},s \nVdash \langle t, (f,b) \rangle \varphi $. In order to $\mathcal{M},s \nVdash \langle t, (f,b) \rangle \varphi$, for every state $v \in S$ such that $s R_\pi v$, $\mathcal{M},v \nVdash \varphi$. Because $f_{ReLo}(t,(f,b)) = \epsilon, s = v$ (i.e., in this execution no other state is reached from $s$). Therefore, $\mathcal{M},v \nVdash \varphi$, contradicting $\mathcal{M},v \Vdash \varphi$.
			
		\end{proof}

\end{lemma}

\subsection{Completeness}


We start by defining the Fisher-Ladner closure of a formula as the set closed by all of its subformulae, following the idea employed in other modal logic works~\cite{Harel2001,Benevides2018} as follows.

\begin{definition}[Fisher-Ladner Closure] \label{def:fl} Let $\Phi$ be a the set of all formulae in \relo.\ The Fischer-Ladner closure of a formula, notation $FL(\varphi)$ is inductively defined as follows:
	
	\begin{itemize}[noitemsep]
		\item $FL\colon \Phi \to 2^\Phi$
		\item $FL_{(f,b)}\colon \{\langle t, (f,b) \rangle \varphi\} \to 2^\Phi$, where $(f,b)$ is a \relo\ program and $\varphi$ a \relo\ formula.
	\end{itemize}
	
	These functions are defined as
	\begin{itemize}[noitemsep]
		\item $FL(p) = \{p\}$, $p$ an atomic proposition;
		\item $FL(\varphi \to \psi) = \{\varphi \to \psi\} \cup FL(\varphi) \cup FL(\psi)$
		\item $FL_{(f,b)}(\langle t, (f,b) \rangle \varphi) = \{\langle t, (f,b) \rangle \varphi\}$
		\item $FL(\langle t, (f,b) \rangle \varphi) = FL_{(f,b)}((\langle t, (f,b) \rangle \varphi) \cup FL(\varphi)$
		\item $FL_{(f,b)}(\langle t, (f,b)^\star \rangle \varphi) = \{\langle t, (f,b)^\star \rangle \varphi\} \cup FL_{(f,b)}(\langle t,(f,b) \rangle \langle t,(f,b)^\star \rangle \varphi)$
		\item $FL(\langle t, (f,b)^\star \rangle \varphi) = FL_{(f,b)}((\langle t, (f,b)^\star \rangle \varphi) \cup FL(\varphi)$
	\end{itemize}
	
\end{definition}

From the definitions above, we prove two lemmas that can be understood as properties that formulae need to satisfy to belong to their Fisher-Ladner closure.

\begin{lemma}\label{lemma:fl1} If $\langle t, (f,b) \rangle \psi \in FL(\varphi)$, then $\psi \in FL(\varphi)$
\end{lemma}

\begin{lemma} \label{lemma:fl2} If $\langle t, (f,b)^\star \rangle \psi \in FL(\varphi)$, then $\langle t, (f,b) \rangle\langle t, (f,b)^\star \rangle \psi \in FL(\varphi)$
\end{lemma}

The proofs for Lemmas~\ref{lemma:fl1} and~\ref{lemma:fl2} are straightforward from Definition~\ref{def:fl}. The following definitions regard the definitions of maximal canonical subsets of \relo\ formulae. We first extend Definition~\ref{def:fl} to a set of formulae $\Gamma$. The Fisher-Ladner closure of a set of formulae $\Gamma$ is $FL(\Gamma) = \bigcup_{\varphi \in \Gamma} FL(\varphi)$. Therefore, $FL(\Gamma)$ is closed under subformulae.
%
For the remainder of this section, we will assume that $\Gamma$ is finite.

\begin{lemma} \label{lemma:finiteFL} If $\Gamma$ is a finite set of formulae, then $FL(\Gamma)$ also is a finite set of formulae
	\begin{proof}
		The proof is standard in literature~\cite{Blackburn2001}. Intuitively, because $FL$ is defined recursively over a set of formulae $\Gamma$ into formulae $\psi$ of a formula $\varphi \in \Gamma$, $\Gamma$ being finite leads to the resulting set of $FL(\Gamma)$ also being finite (at some point, all atomic formulae composing $\varphi$ will have been reached by $FL$).
	\end{proof}
\end{lemma}

\begin{definition}[Atom] \label{def:atom} Let $\Gamma$ be a set of consistent formulae. An atom of $\Gamma$ is a set of formulae  $\Gamma'$ that is a maximal consistent subset of $FL(\Gamma)$. The set of all atoms of $\Gamma$ is defined as $At(\Gamma)$.
\end{definition}

\begin{lemma} \label{lemma:MCSconstruct} Let $\Gamma$ a consistent set of formulae and $\psi$ a \relo\ formula. If $\psi \in FL(\Gamma)$, and $\psi$ is satisfiable then there is an atom of $\Gamma$, $\Gamma'$ where $\psi \in \Gamma'$.
	
	\begin{proof} The proof follows from Lindembaum's lemma. From Lemma~\ref{lemma:finiteFL}, as $FL(\Gamma)$ is a finite set, its elements can be enumerated from $\gamma_1, \gamma_2, \dots, \gamma_n, n = |FL(\Gamma)|$. The first set, $\Gamma'_1$ contains $\psi$ as the starting point of the construction. Then, for $i = 2,\dots, n$, $\Gamma'_i$ is the union of $\Gamma'_{i-1}$ with either $\{\gamma_i\}$ or $\{\neg \gamma_i\}$, respectively whether $\Gamma'_i \cup \{\gamma_i\}$ or 
		$\Gamma'_i \cup \{\neg \gamma_i\}$ is consistent. In the end, we make $\Gamma' = \Gamma'_n$ as it contains the union of all $\Gamma_i, 1 \leq i \leq n$. This is summarized in the following bullets:
		
		\begin{itemize}[noitemsep]
			\item $\Gamma'_1 = \{\psi\}$;
			\item $\displaystyle\begin{aligned}
			\Gamma'_i, = \begin{cases}
			\Gamma'_{i-1} \cup \{\gamma_i\}, \text{ if } \Gamma_{n-1} \cup \{\gamma_n\} \text{ is consistent }\\
			\Gamma'_{i-1} \cup \{\neg \gamma_i\}, \text{ otherwise } \\
			\end{cases}\\
			\end{aligned}$
			for $1 < i < n$; 
			\item	$\Gamma = \bigcup_{i=1}^{n} \Gamma_i$
		\end{itemize}	
		
	\end{proof}
	
\end{lemma}

\begin{definition}[Canonical relations over $\Gamma$]  \label{def:canonicalRelations} Let $\Gamma$ a set of formulae, $A, B$ atoms of $\Gamma$ ($A, B \in At(\Gamma)$), $\Pi$ a \relo\ program and $\langle t, (f,b) \rangle \varphi \in At(\Gamma)$. The canonical relations on $At(\Gamma)$ is defined as $S^{\Gamma}_\Pi$ as follows:
	
	$A S^{\Gamma}_\Pi B \leftrightarrow \bigwedge A \land \langle t, (f,b) \rangle \bigwedge B ) \text{ is consistent }$,
	$A S^{\Gamma}_{\Pi^\star} B \leftrightarrow \bigwedge A \land \langle t, (f,b)^\star \rangle \bigwedge B ) \text{ is consistent }$
\end{definition}

Definition~\ref{def:canonicalRelations} states that the relation between two atoms of $\Gamma$, $A$ and $B$ is done by the conjunction of the formulae in $A$ with all formulae in $B$ which can be accessed from $A$ with a diamond formula, such that this conjunction is also a consistent formula. Intuitively, it states that $A$ and $B$ are related in $S^{\Gamma}_\Pi$ by every formula $\varphi$ of $B$ which conjunction with $A$ by means of a diamond results in a consistent scenario.

The following definition is bound to formalize the canonical version of $\delta$ as the data markup function. 

\begin{definition}[Canonical data markup function $\delta^\Gamma_{c}$]  \label{def:canonicalMarkup}\mbox{}\\ Let $F = \{\langle t_1, (f_1,b_1) \rangle \varphi_1 , \langle t_2, (f_2,b_2) \rangle \varphi_2 , \ldots , \langle t_n, (f_n,b_n) \rangle \varphi_n  \}$ be the set of all diamond formula occurring on an atom $A$ of $\Gamma$. The canonical data markup is defined as $\delta^\Gamma_{c} \colon At(\Gamma) \to T$ as follows:
	
	\begin{itemize}[noitemsep]
		\item The sequence $\{t_1, t_2, \ldots, t_n\} \subseteq \delta(A)$ Therefore, $\{t_1, t_2, \ldots, t_n\} \subseteq \delta^\Gamma_{c}(A)$. Intuitively, this states that all the data flow in the set of formulae must be valid data markups of $A$, which leads to them to also be valid data markups of $\delta^\Gamma_{c}$ following Definition~\ref{def:canonicalRelations}.
		
		\item for all programs $\pi = (f,b) \in \Pi$, $f_{ReLo}((\delta^\Gamma_{c}(A)),(f,b)) \prec \delta^\Gamma_{c}(B) \leftrightarrow A S^{\Gamma}_\Pi B$. 
	\end{itemize}
	
\end{definition}


\begin{definition}[Canonical model] \label{def:canonicalModel} A canonical model over a set of formulae $\Gamma$ is defined as a \relo\ model $\mathcal{M}^\Gamma_c = \langle  At(\Gamma), \Pi ,S^{\Gamma}_\Pi, \delta^\Gamma_{c}, \lambda_c, V^\Gamma_c \rangle $, where:
	
	\begin{itemize}[noitemsep]
		\item $At(\Gamma)$ is the set of states of the canonical model;
		\item $\Pi$ is the model's \relo\ program;
		\item $S^{\Gamma}_\Pi$ are the canonical relations over $\Gamma$;
		\item $\delta^\Gamma_{c}$ is the canonical markup function;
		\item $\lambda_c \colon At(\Gamma) \times \mathcal{N} \rightarrow \mathbbm{R}$;
		\item $V^\Gamma_c \colon At(\Gamma) \times \varphi \to \{true,false\} $, namely $V^\Gamma_c(A,p) = \{A \in At(\Gamma) \mid p \in A\}$;
	\end{itemize}
	
\end{definition}

\begin{lemma} For all programs $\pi = (f,b)$ that compose $\Pi$, $t = \delta^\Gamma_{c}(A)$:
	\begin{enumerate}
		\item If $f_{ReLo}(t,(f,b)) \neq \epsilon$, then $f_{ReLo}(t,(f,b)) \prec \delta^\Gamma_{c}(B)$ iff $A S^{\Gamma}_\Pi B$.
		\item If $f_{ReLo}(t,(f,b)) = \epsilon$, then $(A,B) \notin S^{\Gamma}_\Pi$. 
	\end{enumerate}
	
	\begin{proof}
		The proof for 1. is straightforward from Definition~\ref{def:canonicalMarkup}. The proof for 2. follows from axiom $R$. Because $f_{ReLo}(t,(f,b)) = \epsilon$, no other state is reached from the current state, hence no state $B$ related with $A$ by $R^{\Gamma}_\Pi$ can be reached.
	\end{proof}
\end{lemma}

The following lemma states that canonical models always exists if there is a formula $\langle t, (f,b) \varphi \rangle \in FL(\Gamma)$, a set of formulae $\Gamma$ and a Maximal Consistent Set $A \in At(\Gamma)$. This assures that given the required conditions, a canonical model can always be built.

\begin{lemma}[Existence Lemma for canonical models] \label{lemma:existenceCanonicalModels} Let $A$ be an atom of $At(\Gamma)$ and $\langle t, (f,b) \rangle \varphi \in FL(\Gamma)$. $\langle t, (f,b) \rangle \varphi \in A$ $\iff \exists$ an atom $B \in At(\Gamma)$ such that $A S^{\Gamma}_\Pi B$, $t \prec \delta^\Gamma_{c}(A)$ and $\varphi \in B$.
	\begin{proof} 
		$\Rightarrow$
		Let $A \in At(\Gamma)$ $\langle t, (f,b) \rangle \varphi \in FL(\Gamma)$ and $\langle t, (f,b) \rangle \varphi \in A$ . Because $A \in At(\Gamma)$, from Definition~\ref{def:canonicalMarkup} we have $t \prec \delta^\Gamma_{c} (A)$. From Lemma~\ref{lemma:MCSconstruct} we have that if $\psi \in FL(\Gamma)$ and $\psi$ is consistent, then there is an atom of $\Gamma$, $\Gamma'$ where $\psi \in \Gamma'$. Rewriting $\varphi$ as $(\varphi \land \gamma) \lor (\varphi \land \neg \gamma)$ (a tautology from Propositional Logic), an atom $B \in At(\Gamma)$ can be constructed, because either $\langle t, (f,b) \rangle (\varphi \land \gamma)$ or $\langle t, (f,b) \rangle (\varphi \land \neg \gamma)$ is consistent. Therefore, considering all formulae $\gamma \in FL(\Gamma)$, $B \in At(\Gamma)$ is constructed  with $\varphi \in B$ and $A \land (\langle t, (f,b) \rangle \varphi \bigwedge B$. From Definition~\ref{def:canonicalRelations}, $A S^{\Gamma}_\Pi B$.

		\noindent
		$\Leftarrow$
		Let $A \in At(\Gamma)$ and $\langle t, (f,b) \rangle \varphi \in FL(\Gamma)$. Also, let $B \in At(\Gamma)$, $A S^{\Gamma}_\Pi B$, $t \prec \delta^\Gamma_{c} (A)$, and $\varphi \in B$. As $A S^{\Gamma}_\Pi B$, from Definition~\ref{def:canonicalRelations}, $A S^{\Gamma}_\Pi B \leftrightarrow (A \land \langle t, (f,b) \rangle\bigwedge B),$ $\forall \varphi_i \in B$ is consistent. From $\varphi \in B$, $(A \land \langle t, (f,b) \rangle \varphi)$ is also consistent. As $A \in At(\Gamma)$ and $\langle t, (f,b) \varphi \in FL(\Gamma)$, by Definition~\ref{def:atom}, as $A$ is maximal, then $\langle t, (f,b) \rangle \varphi \in A$.
		
	\end{proof}
\end{lemma}

The following lemma formalizes the truth notion for a canonical model $\mathcal{M}^{\Gamma}_c$, given a state $s$ and a formula $\varphi$. It formalizes the semantic notion for canonical models in \relo.

\begin{lemma}[Truth Lemma] Let $\mathcal{M}^\Gamma_c = \langle  At(\Gamma), \Pi ,S^{\Gamma}_\Pi, \delta^\Gamma_{c}, \lambda, V^\Gamma_c \rangle $ be a canonical model over a formula $\gamma$. Then, for every state $A \in At(\Gamma)$ and every formula $\varphi \in FL(\gamma)$:
$\mathcal{M}^\Gamma_c, A \Vdash \varphi \iff \varphi \in A$.
	
	\begin{proof} The proof proceeds by induction over the structure of $\varphi$.
		\begin{itemize}[noitemsep]
			\item Induction basis: suppose $\varphi$ is a proposition $p$. Therefore, $\mathcal{M}^\Gamma_c, A \Vdash p$. From Definition~\ref{def:canonicalModel}, $\mathcal{M}^{\Gamma}_c$'s valuation function is $V^\Gamma_c(p) = \{A \in At(\Gamma) \mid p \in A\}$. Therefore, $p \in A$.
			
			\item Induction Hypothesis: Suppose $\varphi$ is a non atomic formula $\psi$. Then, $\mathcal{M}^\Gamma_c, A \Vdash \psi \iff \psi \in A$, $\psi$ a strict subformula of $\varphi$.
			
			\item Inductive step: Let us prove it holds for the following cases (we ommit propositional operators):
			
			\begin{itemize}[noitemsep]
				
				
				
				\item Case $\varphi = \langle t, (f,b) \rangle \phi$. Then, $\mathcal{M}^\Gamma_c, A \Vdash \langle t, (f,b) \rangle \phi \iff \langle t, (f,b) \rangle \phi \in A$:
				\mbox{}\\ \noindent
				$\Rightarrow$
				Let $\mathcal{M}^\Gamma_c, A \Vdash \langle t, (f,b) \rangle \phi$. From Definition~\ref{def:canonicalRelations}, there is a state $B$ where $A S^{\Gamma}_\Pi B$ and $\phi \in B$. By Lemma~\ref{lemma:existenceCanonicalModels}, $\langle t, (f,b) \rangle \phi \in A$. Therefore, it holds.
				\mbox{}\\ \noindent
				$\Leftarrow$	
				Let	$\mathcal{M}^\Gamma_c, A \nVdash \langle t, (f,b) \rangle \phi$. From Definition~\ref{def:canonicalModel}'s valuation function $V^\Gamma_c$ and Lemma~\ref{lemma:MCSconstruct}, we have $\mathcal{M}^\Gamma_c, A \Vdash \neg\langle t, (f,b) \rangle \phi$. Therefore, for every $B$ where $A S^\Gamma_\Pi B, \mathcal{M}^\Gamma_c, B \Vdash \neg \phi$. From the induction hypothesis, $\phi \notin B$. Hence, From Lemma~\ref{lemma:existenceCanonicalModels}, $\langle t, (f,b) \rangle \phi \notin A$.
				
				\item Case $\varphi = \langle t, (f,b)^\star \rangle \phi$. Then, $\mathcal{M}^\Gamma_c, A \Vdash \langle t, (f,b)^\star \rangle \phi \iff \langle t, (f,b)^\star \rangle \phi \in A$:
				\mbox{}\\ \noindent
				$\Rightarrow$
				Let $\mathcal{M}^\Gamma_c, A \Vdash \langle t, (f,b)^\star \rangle \phi$. From Definition~\ref{def:canonicalRelations}, there is a state $B$ where $A S^{\Gamma}_{\Pi^\star} B$ and $\phi \in B$. By Lemma~\ref{lemma:existenceCanonicalModels}, $\langle t, (f,b)^\star \rangle \phi \in A$. Therefore, it holds.
				\mbox{}\\\noindent
				$\Leftarrow$	
				Let	$\mathcal{M}^\Gamma_c, A \nVdash \langle t, (f,b)^\star \rangle \phi$. From Definition~\ref{def:canonicalModel}'s valuation function $V^\Gamma_c$ and Lemma~\ref{lemma:MCSconstruct}, we have $\mathcal{M}^\Gamma_c, A \Vdash \neg\langle t, (f,b)^\star \rangle \phi$. Therefore, for every $B$ where $A S^\Gamma_{\Pi^\star} B, \mathcal{M}^\Gamma_c, B \Vdash \neg \phi$. From the induction hypothesis, $\phi \notin B$. Hence, From Lemma~\ref{lemma:existenceCanonicalModels}, $\langle t, (f,b)^\star \rangle \phi \notin A$.
				
			\end{itemize}
		\end{itemize}
	\end{proof}
	
\end{lemma}

We proceed by formalizing the following lemma, which is bound to show that the properties that define $\star$ for regular \relo\ models also holds in \relo\ canonical models.

\begin{lemma} Let $A, B \in At(\Gamma)$ and $\Pi$ a \relo\ program. If $A S_{\Pi^\star} B$ then $A S_\Pi^\star B$ \label{lemma:starCanonicalRelations}
	
	\begin{proof}
		Suppose $A S_{\Pi^\star} B$. Define $C = \lbrace C' \in At(\Gamma) \mid A S_\Pi^\star C \rbrace$ as the set of all atoms $C'$ which $A$ reaches by means of $S_{\Pi^\star}$. We will show that $B \in C$. Let $C_c$ be the maximal consistent set obtained by means of Lemma~\ref{lemma:MCSconstruct}, $C_c = \{ \bigwedge C_1 \lor C_2 \lor \dots \bigwedge C_n\}$, where the conjunction of each $C_i$ is consistent, and each $C_i$ is a maximal consistent set. Also, define $t = \delta_c^\Gamma(C_c)$ as the canonical markup of $C_c$.
		
		Note that $C_c \land \langle t, (f,b) \rangle \neg C_c$ is inconsistent: if it was consistent, then for some $D \in At(\Gamma)$ which $A$ cannot reach, $C_c \land \langle t, (f,b) \rangle \bigwedge D$ would be consistent, which leads to $\bigwedge C_1 \lor C_2 \lor \dots \lor C_i \lor \langle t, (f,b) \rangle \bigwedge D$ also being consistent, for some $C_i$. By the definition of $C_c$, this means that $D \in C$ but that is not the case (because $D \in C_c$ contradicts $D$ not being reached from $A$ and consequently $C_c$'s definition, as $D \in C_c$ leads to D being reachable from $A$). Following a similar reasoning, $\bigwedge A \land \langle t, (f,b) \rangle C_c$ is also inconsistent and therefore its negation, $\bigwedge \neg (A \land \langle t, (f,b) \rangle C_c)$ is consistent, which can be rewritten as $\bigwedge A \rightarrow \lbrack t, (f,b) \rbrack C_c$.
		
		
		Because $C_c \land \langle t, (f,b) \rangle \neg C_c$ is inconsistent, its negation $\neg (C_c \land \langle t, (f,b) \rangle \neg C_c)$ is valid, which can be rewritten to $\vdash C_c \rightarrow \lbrack t, (f,b) \rbrack C_c$ (I). Therefore, by applying generalization we have $\vdash \lbrack t, (f,b)^\star \rbrack (C_c \rightarrow \lbrack t, (f,b) \rbrack C_c)$. By axiom \textbf{(It)}, we derive $\vdash \lbrack t, (f,b) \rbrack C_c \rightarrow \lbrack t, (f,b)^\star \rbrack C_c$ (II). By rewriting (II) in (I) we derive $C_c \rightarrow \lbrack t, (f,b)^\star \rbrack C_c$. As $\bigwedge A \rightarrow \lbrack t, (f,b) \rbrack C_c$ is valid, from (II) $\bigwedge A \rightarrow \lbrack t, (f,b)^\star \rbrack C_c$ also is valid. From the hypothesis $A S_{\pi^\star} B$ and $C_c$'s definition, $\bigwedge A \land \langle t, (f,b)^\star \rangle B$ and $\bigwedge B \land C_c$ are consistent (the latter from $C_c$'s definition). Then, there is a $C_i \in C_c$ such that $\bigwedge B \land \bigwedge C$ is consistent. But because each $C_i$ is a maximal consistent set, it is the case that $B = C_i$, which by the definition of $C_c$ leads to $A S_\Pi^\star B$.
		
	\end{proof}
	
\end{lemma}

\begin{definition}[Proper Canonical Model] The proper canonical model over a set of formulae $\Gamma$ is defined as a tuple $\langle At(\Gamma), \Pi, R^{\Gamma}_\Pi, \delta^{\Gamma}_\Pi, \lambda_c, V^{\Gamma}_\Pi  \rangle$ as follows: \label{def:propercanonicalmodel}
	
	\begin{itemize}[noitemsep]
		\item $At(\Gamma)$ as the set of atoms of $\Gamma$;
		\item $\Pi$ as the \relo\ program;
		\item The relation $R$ of a \relo\ program $\Pi$ is inductively defined as:
		\begin{itemize}[noitemsep]
			\item $R_\pi = S_\pi$ for each canonical program $\pi$;
			\item $R^\Gamma_{\Pi^\star} = (R^\Gamma_{\Pi})^{\star}$;
			\item $\Pi = \pi_1 \odot \pi_2 \odot \dots \odot \pi_n$ a \relo\ program, $R_\Pi \subseteq S \times S$ as follows:
			\begin{itemize}[noitemsep]
				\item $R_{\pi_i} =	\lbrace uR_{\pi_i}v \mid f_{ReLo}(t,\pi_i) \prec \delta(v) \rbrace$, $t \prec \delta(u)$ and $\pi_i$ is any combination of any atomic programs which is a subprogram of $\Pi$.
			\end{itemize}
			%
			
		\end{itemize}
		\item $\delta^{\Gamma}_\Pi$ as the canonical markup function;
		\item $\lambda_c \colon At(\Gamma) \times \mathcal{N} \rightarrow \mathbbm{R}$;
		\item $V^\Gamma_c(A,p) = \{A \in At(\Gamma) \mid p \in A\}$ as the canonical valuation introduced by Definition~\ref{def:canonicalModel}.
		
	\end{itemize}
\end{definition}

\begin{lemma} Every canonical model for $\Pi$ has a corresponding proper canonical model: for all programs $\Pi$, $S^{\Gamma}_\Pi \subseteq R^{\Gamma}_\Pi$ \label{lemma:SsubseteqR}
	
	%
	
	\begin{proof} The proof proceeds by induction on $\Pi$'s length
		\begin{itemize}[noitemsep]
			\item For basic programs $\pi$, it follows from Definition~\ref{def:propercanonicalmodel}:  
			\item $\Pi^\star$: From Definition~\ref{def:reloFrame}, $R_{\pi^\star} = R^\star_\pi$. By the induction hypothesis, $S^{\Gamma}_{\Pi} \subseteq R^{\Gamma}_{\Pi}$, also from the definition of RTC, we have that if $(S^{\Gamma}_{\Pi}) \subseteq (R^{\Gamma}_{\Pi})$, then $(S^{\Gamma}_{\Pi})^\star \subseteq (R^{\Gamma}_{\pi})^\star$ (i). From Lemma~\ref{lemma:starCanonicalRelations}, $S^{\Gamma}_{\Pi^\star} \subseteq (S^{\Gamma}_\Pi)^\star$, which leads to $(S^{\Gamma}_\Pi)^\star\subseteq (R^{\Gamma}_{\Pi})^\star$ by (i). Finally, $(R^{\Gamma}_{\Pi})^\star = (R^{\Gamma}_{\Pi^\star})$. Hence, $(S^{\Gamma}_{\Pi^\star}) \subseteq (R^{\Gamma}_{\Pi^\star})$
			
			%
		\end{itemize}
	\end{proof}
\end{lemma}

\begin{lemma}[Existence Lemma for Proper Canonical Models] \label{lemma:existenceLemmaProperCanonicalModel}Let $A \in At(\Gamma)$ and $\langle t, (f,b) \rangle\varphi \in FL(\Gamma)$. Then,
	$\langle t, (f,b) \rangle \varphi \in A \leftrightarrow \text{exists } B \in At(\Gamma), A R^\Gamma_\Pi B, t \prec \delta^{\Gamma}_c(A) \text{ and } \varphi \in B.$
	\begin{proof}
		$\Rightarrow$
		Let $\langle t, (f,b) \rangle \varphi \in A$. From Lemma~\ref{lemma:existenceCanonicalModels} (Existence Lemma for canonical models), there is an atom $B \in At(\Gamma)$ where $ A S^\Gamma_{\Pi} B$, $t \prec \delta^\Gamma_c (A)$ and $\varphi \in B$. From Lemma~\ref{lemma:SsubseteqR}, $S^{\Gamma}_\Pi \subseteq R^{\Gamma}_\Pi$. Therefore, there is an atom $B \in At(\Gamma)$ where $ A R^\Gamma_{\Pi} B$, $t \prec \delta^\Gamma_c (A)$ and $\varphi \in B$.
		
		\noindent
		$\Leftarrow$
		Let $B$ an atom, $B \in At(\Gamma), A R_\Pi B, t \prec \delta^{\Gamma}_c(A) \text{ and } \varphi \in B$. The proof follows by induction on the program $\Pi = (f,b)$ as follows:
		\begin{itemize}[noitemsep]
			\item a canonical program $\pi_i$: this case is straightforward as from Definition~\ref{def:propercanonicalmodel}, $S_{\pi_i} = R_{\pi_i}$, and consequently $A S_{\pi_i} B, t \prec \delta^{\Gamma}_c (A)$ and (i) $\varphi \in B$. From Lemma~\ref{lemma:existenceCanonicalModels} and (i), $\langle t, (f,b) \rangle \varphi \in A$.
			
			\item $\Pi^\star$: from Definition~\ref{def:propercanonicalmodel}, $R_{\Pi^\star} = R^\star_{\Pi}$. Then, let $B \in At(\Gamma), A R_{\Pi^\star} B, t \prec \delta^{\Gamma}_c(A) \text{ and } \varphi \in B$. This means that there is a finite nondeterministic number $n$ where 
			$A R_{\Pi^\star} B = A R_{\Pi} A_1 R_{\Pi} A_2 \dots R_{\Pi} A_n$, where $A_n = B$. The proof proceeds by induction on $n$:
			\begin{itemize}[noitemsep]
				\item $n = 1$: $A R_{\Pi} B$ and $\varphi \in B$. Therefore, from Lemma~\ref{lemma:existenceCanonicalModels},$\langle t, (f,b) \rangle \varphi \in A$. From axiom Rec, one may derive $\Vdash \langle t, (f,b) \rangle \varphi \rightarrow \langle t, (f,b)^\star \rangle \varphi$. By the definition of FL and $A$'s maximality (as it is an atom of $\Gamma)$ $\langle t, (f,b)^\star \rangle \varphi \in A$.
				
				\item $n > 1$: From the previous proof step and the induction hypothesis, $\langle t, (f,b)^\star \rangle \in A_2$ and $\langle t, (f,b) \rangle \langle t, (f,b)^\star \rangle \in A_1$. From axiom Rec, one can derive \\$\Vdash \langle t, (f,b) \rangle \langle t, (f,b)^\star \rangle \varphi \rightarrow \langle t, (f,b)^\star \rangle \varphi$. By the definition of $FL$, and $A$'s maximality (as it is an atom of $\Gamma)$, $\langle t, (f,b)^\star \rangle \varphi \in A$.
			\end{itemize}
		\end{itemize}
		
	\end{proof}
\end{lemma}

\begin{lemma}[Truth Lemma for Proper Canonical Models] Let $\mathcal{M}^\Gamma_c = \langle At(\Gamma), \Pi, R^{\Gamma}_\Pi, \delta^{\Gamma}_\Pi, \lambda_c, V^{\Gamma}_\Pi  \rangle$ a proper canonical model constructed over a formula $\gamma$. For all atoms $A$ and all $\varphi \in FL(\gamma).$ $\mathcal{M}, A \Vdash \varphi \leftrightarrow \varphi \in A.$ \label{lemma:truthLemmaProperCanonicalModel}
	
	\begin{proof}
		The proof proceeds by induction over $\varphi$.
		\begin{itemize}[noitemsep]
			\item Induction basis: $\varphi$ is a proposition p. Therefore, $\mathcal{M}^\Gamma_c, A \Vdash p$ holds from Definition~\ref{def:propercanonicalmodel} as $V^\Gamma_c(p) = \{A \in At(\Gamma) \mid p \in A\}$.
			
			\item Induction hypothesis: suppose $\varphi$ is a non atomic formula $\psi$. Then,  $\mathcal{M}, A \Vdash \varphi \iff \varphi \in A$, $\psi$ a strict subformula of $\varphi$.
			
			\item Inductive step: let us prove it holds for the following cases (we show only for modal cases):
			
			\begin{itemize}[noitemsep]
				
				
				
				\item Case $\varphi = \langle t, (f,b) \rangle \phi$. Then, $\mathcal{M}^\Gamma_c, A \Vdash \langle t, (f,b) \rangle \phi \iff \langle t, (f,b) \rangle \phi \in A$:
				\mbox{}\\ \noindent
				$\Rightarrow$
				Let $\mathcal{M}^\Gamma_c, A \Vdash \langle t, (f,b) \rangle \phi$. From Definition~\ref{def:canonicalRelations}, there is an atom $B$ where $A S^{\Gamma}_\Pi B$ and $\phi \in B$. By Lemma~\ref{lemma:existenceLemmaProperCanonicalModel}, $\langle t, (f,b) \rangle \phi \in A$. Therefore, it holds.
				\mbox{}\\ \noindent
				$\Leftarrow$		
				Let	$\mathcal{M}^\Gamma_c, A \nVdash \langle t, (f,b) \rangle \phi$. From Definition~\ref{def:canonicalModel}'s valuation function $V^\Gamma_c$ and Lemma~\ref{lemma:MCSconstruct}, we have $\mathcal{M}^\Gamma_c, A \Vdash \neg\langle t, (f,b) \rangle \phi$. Therefore, for every $B$ where $A S^\Gamma_\Pi B, \mathcal{M}^\Gamma_c \Vdash \neg \phi$. From the induction hypothesis, $\phi \notin B$. Hence, from Lemma~\ref{lemma:existenceLemmaProperCanonicalModel} $\langle t, (f,b) \rangle \phi \notin A$.			
				%
				\item Case $\varphi = \langle t, (f,b)^\star \rangle \phi$. Then, $\mathcal{M}^\Gamma_c, A \Vdash \langle t, (f,b)^\star \rangle \phi \iff \langle t, (f,b)^\star \rangle \phi \in A$:
				\mbox{}\\ \noindent
				$\Rightarrow$
				Let $\mathcal{M}^\Gamma_c, A \Vdash \langle t, (f,b)^\star \rangle \phi$. From Definition~\ref{def:canonicalRelations}, there is a state $B$ where $A S^{\Gamma}_{\Pi^\star} B$ and $\phi \in B$. By Lemma~\ref{lemma:existenceCanonicalModels}, $\langle t, (f,b)^\star \rangle \phi \in A$. Therefore, it holds.
				\mbox{}\\\noindent
				$\Leftarrow$	
				Let	$\mathcal{M}^\Gamma_c, A \nVdash \langle t, (f,b)^\star \rangle \phi$. From Definition~\ref{def:canonicalModel}'s valuation function $V^\Gamma_c$ and Lemma~\ref{lemma:MCSconstruct}, we have $\mathcal{M}^\Gamma_c, A \Vdash \neg\langle t, (f,b)^\star \rangle \phi$. Therefore, for every $B$ where $A S^\Gamma_{\Pi^\star} B, \mathcal{M}^\Gamma_c, B \Vdash \neg \phi$. From the induction hypothesis, $\phi \notin B$. Hence, From Lemma~\ref{lemma:existenceCanonicalModels}, $\langle t, (f,b)^\star \rangle \phi \notin A$.
			\end{itemize}
		\end{itemize}
	\end{proof}  
	
\end{lemma}

\begin{theorem}[Completeness of \relo] \label{lemma:reloCompleteness} 
	\begin{proof}
		For every consistent formula $A$, a canonical model $\mathcal{M}$ can be constructed. From Lemma~\ref{lemma:MCSconstruct}, there is an atom $A' \in At(A)$ with $A \in A'$, and from Lemma~\ref{lemma:truthLemmaProperCanonicalModel}, $\mathcal{M}, A' \Vdash A$. Therefore, \relo's modal system is complete with respect to the class of proper canonical models as Definition~\ref{def:propercanonicalmodel} proposes.
	\end{proof}
\end{theorem}

\section{Conclusions and Further Work} \label{sec:conclusions}

Reo is a widely used tool to model new systems out of the coordination of already existing pieces of software. It has been used in a variety of domains, drawing the attention of researchers from different locations around the world. This has resulted in Reo having many formal semantics proposed, each one employing different formalisms: operational, co-algebraic, and coloring semantics are some of the types of semantics proposed for Reo.

This work extends \relo, a dynamic logic to reason about Reo models. We have discussed its core definitions, syntax, semantic notion, providing soundness and completeness proofs for it.
\relo\ naturally subsumes the notion of Reo programs and models in its syntax and semantics, and implementing its core concepts in Coq enables the usage of Coq's proof apparatus to reason over Reo models with \relo.

Future work may consider the integration of the current implementation of \relo\ with ReoXplore\footnote{\url{https://github.com/frame-lab/ReoXplore2}}, a platform conceived to reason about Reo models, and extensions to other Reo semantics. Investigations and the development of calculi for \relo\ are also considered for future work.


\bibliographystyle{eptcs}
\bibliography{ref} 



\end{document}